\documentstyle[prcrc,11pt,fleqn,epsf]{article}

\title{Sunyaev-Zel'dovich Effect as a Cosmological Probe}
\author{Asantha R. Cooray\thanks{Department of 
Astronomy and Astrophysics, University of Chicago, Chicago IL 60637.}, 
John E. Carlstrom\thanks{Department of 
Astronomy and Astrophysics, University of Chicago, Chicago IL 60637.}, 
Marshall Joy\thanks{Space Science Laboratory, NASA Marshall Space Flight Center, Huntsville AL 35812.},
Laura Grego\thanks{Department of Astronomy and Astrophysics, University of 
Chicago, Chicago IL 60637.}, 
William L. Holzapfel\thanks{Department of 
Astronomy and Astrophysics, University of Chicago, Chicago IL 60637.}
Sandeep K. Patel\thanks{Department of Physics, University of Alabama in Huntsville, Huntsville AL 35899.}
}

\begin{document}
\maketitle

\begin{abstract}

We review recent results of Sunyaev-Zel'dovich effect (SZE) observations
toward galaxy clusters. Using cm-wave receivers mounted on the OVRO and
BIMA mm-wave arrays we have obtained high signal to noise images of the
effect for more than 20 clusters.  We present current estimates of the
Hubble constant and cosmological parameters and discuss the potential of
conducting statistical studies with large SZE cluster samples.

\end{abstract}

\section{Introduction}

Over the last few years there has been a tremendous increase in 
the study of galaxy clusters 
as cosmological probes, initially through the
use of X-ray emission observations, and in recent years,
through the use of Sunyaev-Zel'dovich effect (SZE).
Briefly, the SZE is
a distortion of the cosmic microwave background
(CMB) radiation by inverse-Compton scattering of 
thermal electrons within the hot intercluster medium
(Sunyaev \& Zel'dovich 1980, see Birkinshaw 1998 for a recent review). 
The change in the CMB brightness temperature observed
is:
\begin{equation}
\frac{\Delta T}{T_{\rm CMB}} = \left[ \frac{x (e^{x}+1)}{e^{x}-1} -4 \right]
\int \left(\frac{k_B T_e}{m_e c^2}\right) n_e \sigma_T dl,
\end{equation}
where $x = h \nu/k_B T_{\rm CMB}$, and $n_e$, $T_e$ and $\sigma_T$ are the 
electron density, electron temperature and the cross section for Thomson 
scattering. The integral is performed along the line of sight through the
cluster. 

The other important
observable of the hot intercluster gas 
is the thermal Bremsstrahlung
X-ray emission, whose surface brightness $S_X$ can be written 
as:
\begin{equation}
S_X = \frac{1}{4 \pi (1+z)^3} \int n ^{2}_{e} \Lambda_e dl,
\end{equation}
where $z$ is the redshift and $\Lambda_e(\Delta E,T_e)$ is the
X-ray spectral emissivity of the cluster gas due to thermal
Bremsstrahlung within a certain energy band $\Delta E$.
By combining the intensity of the SZE and the X-ray emission observations,
and knowing the cluster gas temperature $T_e$,
the angular diameter distance to the cluster can be derived due to the
different dependence of the X-ray emission and SZE
on the electron density, $n_e$.
Combining such distance measurements with redshift
allows a determination of the 
Hubble constant, H$_0$, as a function of certain cosmological
parameters (e.g., Hughes \& Birkinshaw 1998a).
If distance measurements for a sample of
clusters exist, then the angular diameter distance with redshift 
relation can be used to put constraints on the cosmological models,
similar to current supernovae constraints at high redshift. 

\section{Interferometric Observations of the SZ Effect}

We have imaged the SZE by outfitting the OVRO and BIMA mm-wave
arrays with low-noise cm-wave receivers. 
One of the key advantages of our system
is the ability to use interferometric techniques to produce
2-dimensional images of the SZE with sensitivity to
large angular scales (up to $2.5'$).
The system as installed
at OVRO and the first images obtained are discussed in 
Carlstrom {\it et al.} (1996). In Cooray {\it et al.} (1998a),
we presented the observed cluster sample at OVRO and BIMA during the
summers of 1995 to 1997 and detections of
radio sources in galaxy clusters at 28.5 GHz. One of the
main problems of SZE observations at cm-wavelengths is the presence of 
bright radio sources towards galaxy clusters, and catalogs of such sources at 
are important for future SZE and CMB anisotropy observations. 

In Figure 1 we present images of the galaxy cluster A2218. 
The third panel shows our SZ image. The cluster
was observed for 60 hours, producing this map with a rms noise of 15 $\mu$K. 
Currently, we have imaged the SZE with high signal-to-noise
($>$ 20) in $\sim$ 20 clusters from z $\sim$ 0.14 to 0.83. Over 80\%
of this sample is scheduled to be observed with AXAF during the GTO phase.
When combined, the SZE and the AXAF
X-ray emission data will allow the determination of the Hubble constant, and
constraints on the cosmological parameters based on the angular diameter
distance relation 
with redshift, at a level comparable to the present SNIa constraints
on cosmological parameters.

\begin{figure}[t]
\vbox to1.2in{\rule{0pt}{1in}}
\includegraphics{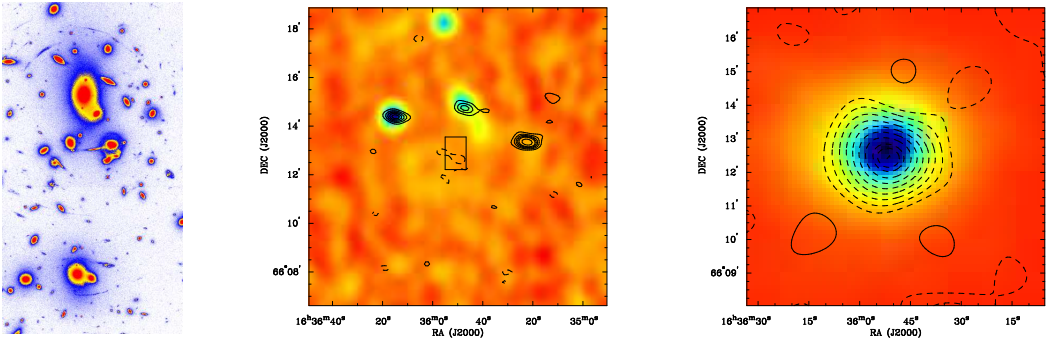}
\includegraphics{figure2.eps}
\includegraphics{figure3.eps}
\caption{Three views of the cluster A2218: (a) The HST image of
the central core region (Kneib et al. 1996) (b) The BIMA 28.5 GHz 
naturally weighted contours with VLA D-array NVSS observations 
in the background. The small rectangle roughly indicates the region of
the HST image. (c) The detected SZE, 
after accounting for the bright radio sources. The background
of this map is the 
ROSAT PSPC image, smoothed with a 20$''$ FWHM Gaussian.}

\end{figure}

\section{Cosmological Parameter Constraints}

\subsection{$H_0$}

\begin{table}[hbt]
\newlength{\digitwidth} \settowidth{\digitwidth}{\rm 0}
\catcode`?=\active \def?{\kern\digitwidth}
\caption{SZ Effect/X-ray Angular Diameter Distance and $H_0$ Measurements}
\begin{tabular}{lllcl}
\hline
Cluster & Redshift & D$_{\rm A}$ (Mpc) &$H_0$ (km s$^{-1}$ Mpc$^{-1}$)& Reference \\
\hline
A2256 & 0.0581 & $231^{+82}_{-54}$ & 68$^{+21}_{-18}$ & Myers {\it et al.} 1997\\
A478 & 0.0881 & $747^{+539}_{-264}$& 30$^{+17}_{-13}$ & Myers {\it et al.} 1997\\
A2142 & 0.0899 & $512^{+797}_{-265}$ & 46$^{+41}_{-28}$ & Myers {\it et al.} 1997 \\
A1413 & 0.143 & $743^{+348}_{-222}$ & 44$^{+20}_{-15}$ & Saunders 1996 \\
A2218 & 0.171 & $678^{+432}_{-190}$ & 59 $\pm$ 23 & Birkinshaw \& Hughes 1994 \\
A2218 & 0.171 & $1176^{+823}_{-376}$ & 34$^{+18}_{-16}$ & Jones 1995\\
A2218 & 0.171 & $1050 \pm 230$     &  38 $\pm$ 15  & Patel {\it et al.} 1998 \\
A665 & 0.182 & $911^{+235}_{-486}$ & 46 $\pm$ 16 & Hughes \& Birkinshaw 1998b \\
A665 & 0.182 & $939^{+260}_{-495}$ & 48$^{+19}_{-16}$ & Cooray {\it et al.} 1998b \\
A2163 & 0.201 & $778^{+475}_{-313}$ & 58$^{+39}_{-22}$ & Holzapfel {\it et al.} 1997\\
Cl0016+16 & 0.5455 & $1713^{+803}_{-562}$ & 47$^{+23}_{-15}$ & Hughes \& Birkinshaw 1998a\\

\hline
\multicolumn{5}{@{}p{120mm}}{$H_0$ is calculated assuming
$\Omega_m=0.2$ and $\Omega_{\Lambda}=0$.}
\end{tabular}
\end{table}

Table 1 lists published Hubble constant measurements that have been obtained
by combining SZE and X-ray emission observations (see Hughes 1997 for
further details). 
In recent years, several studies have questioned
the reliability of $H_0$ measurements based on SZ/X-ray route. This is
primarily due to various 
systematic effects involved with this method,  which include
the nonisothermality of the electron temperature for
cluster gas,
gas clumping, asphericity of the cluster gas
distribution, and radio source contamination and
gravitational lensing effects (see Birkinshaw 1998). 
It is likely that deep AXAF observations
will produce reliable cluster electron temperature profiles and
constrain the amount of gas clumping. 
Systematic changes in $H_0$ due to aspherical gas
distribution can be treated in a statistical manner
for a large sample of clusters.

\subsection{Gas Fraction and $\Omega_m$}

The SZE is a measurement of the integrated gas (baryonic) 
mass along the line of slight through the cluster. The total (including non-baryonic)
mass of a cluster can be derived based on three methods:
gas temperature, gravitational (strong \& weak) lensing, 
and velocity dispersion measurements.
The baryonic mass fraction, when compared to the primordial nucleoynthesis
determined value for $\Omega_b$ allows constraints on $\Omega_m$
assuming the cluster baryonic fraction
is the same as $\Omega_b/\Omega_m$ (based on 
hierarchical clustering models, where clusters
represent the composition of the universe).
The present limits on $\Omega_m$ are:
$\Omega_m h < 0.3$ (Grego {\it et al.} 1998, Myers {\it et al.} 1997), based
on SZE measurements, 
and $\Omega_m h^{2/3} < 0.28$ (Evrard, this proceedings), based
on X-ray measurements.

\subsection{D$_A$ and $q_0$, $\Omega_m$, $\Omega_{\Lambda}$}

The angular diameter distance relation with redshift is dependent on the
cosmological parameters. Thus, if distance measurements exist
out to high redshift, one can use the angular diameter distance with redshift 
to constrain the cosmological parameters. Present
SZ/X-ray $D_A$ measurements (in Table 1) do not allow reliable constraints
on the $\Omega_m-\Omega_{\Lambda}$ plane; the data in Table 1
are consistent 
with $q_0 > -0.73$ (90\% formal confidence).


\section{Scaling Relations as Cosmological Tools}

One of the well known facts about the SZE is that it is independent
of redshift, allowing a probe of the distant universe. Given that large
area SZE surveys and {\it PLANCK} will detect a large number of
high redshift clusters, it is necessary that techniques independent
of X-ray observations be considered.
The temperature change due to the SZE is expected to relate
to the X-ray luminosity, and 
the expected relation takes the form of $\Delta T_{\rm SZ} \propto
L_{\rm bol}^{\alpha}$  with $\alpha =0.6$ to $0.7$ depending on the
exact form of the $L_{\rm bol}-T_e$ relation. The present SZ data, 
suggest
\begin{equation}
\Delta T_{\rm SZ} = -(0.46 \pm 0.12) \left(\frac{L_{\rm bol}}{10^{45}}\right)^{0.60 \pm 0.18}\, {\rm mK},
\end{equation}
where the uncertainty is the $1-\sigma$ statistical error.
The use of such relations, say in a Press-Schechter formalism, can
be helpful to constrain the cosmological parameters based on the SZE 
observations, with the relations normalized based on the X-ray data
 of the local universe.
Another important scaling relation would be between the SZE and
the cluster electron temperature. Since the
 SZE measures the cluster gas mass
and cluster temperature measures the total mass, this relation
will probe the gas mass fraction in galaxy clusters. The redshift 
evolution of $\Delta T_{\rm SZ} - T_{\rm e0}$ 
relation can be used to derive cosmological parameters, similar
to what is done in Cooray (1998) for X-ray/lensing data (see also
Danos \& Pen 1998).

\begin{small}

\end{small}

\end{document}